\documentstyle[preprint,aps]{revtex}

\begin{document}
\draft
\author{Bo-Bo Wang\thanks{%
E-mail: bbwang@mail.263.net.cn} and Liao Liu\thanks{%
E-mail: liaoliu@bnu.edu.cn}}
\address{Department of Physics, Beijing Normal University, Beijing 100875, China.}
\title{The statistic thermodynamical entropy of de Sitter space and it's 1-loop
correction}
\maketitle

\begin{abstract}
From the partition function of canonical ensemble we derive the entropy of
the de Sitter space by anti-Wick rotation. And then from the one-loop bubble 
$S^2\times S^2$ created from vacuum fluctuation in de Sitter background
space, we obtain the one-loop quantum correction to the entropy of de Sitter
space.
\end{abstract}

\pacs{PACS numbers: 04.70.Dy.}

\section{The entropy of the de Sitter space}

Gibbons and Hawking was the first one who argued that the de Sitter space
has entropy which equals to the fourth of the area of cosmological horizon
of the de Sitter space\cite{G-H}. However a derivation based on canonical
ensemble in quantum statistic thermodynamics is wanted.

As in order to cancel the coordinate singularity and to reveal the imaginary
period of time variable of the static de Sitter space 
\begin{equation}
ds^2=-\left( 1-\frac{\Lambda r^2}3\right) dt^2+\left( 1-\frac{\Lambda r^2}3%
\right) ^{-1}dr^2+r^2d\Omega ^2,  \label{metric}
\end{equation}
where $\Lambda $ is the cosmological constant, $d\Omega ^2=d\theta ^2+\sin
^2\theta d\varphi ^2$ is the unit 2-sphere. One should introduce a
Kruskal-like coordinate\cite{G-H} $\left( U,V,\theta ,\varphi \right) $ by 
\begin{equation}
r=\sqrt{\frac 3\Lambda }\frac{UV+1}{1-UV},  \label{B1}
\end{equation}
\begin{equation}
\exp \left( 2\sqrt{\frac \Lambda 3}t\right) =-VU^{-1},  \label{B2}
\end{equation}
then Eq.$\left( \ref{metric}\right) $ becomes 
\begin{equation}
ds^2=3\Lambda ^{-1}\left( UV-1\right) ^{-2}\left[ -4dUdV+\left( UV+1\right)
^2d\Omega ^2\right] .  \label{metric2}
\end{equation}
From Eqs.$\left( \ref{B1}\right) $ and $\left( \ref{B2}\right) $, the
space-time has clearly an imaginary period of 
\begin{equation}
\beta =2\pi \sqrt{\frac 3\Lambda }.
\end{equation}
As is known, for any system including the gravity, if time has an imaginary
period then by using Schr\"{o}dinger picture, the partition function of the
system will be the partition function of the canonical ensemble\cite{Hawking}%
, i.e. 
\begin{equation}
Z=\sum_n\exp \left( -E_n\beta \right) =\int D\left[ g\right] \exp \left( -%
\hat{I}\left[ g\right] \right)  \label{Z}
\end{equation}
and 
\begin{equation}
\ln Z=-\hat{I},  \label{p}
\end{equation}
where the summation is over the different quantum state, $\hat{I}$ is the
Euclidean action, Eq.$\left( \ref{p}\right) $ holds in saddle point
approximation. From 
\begin{equation}
S=-\sum_nP_n\ln P_n,
\end{equation}
and 
\begin{equation}
P_n=Z^{-1}\exp \left( -\beta E_n\right)
\end{equation}
we get the entropy $S$ and the energy $\left\langle E\right\rangle $ of the
gravitating system 
\begin{equation}
\left\langle E\right\rangle =-\frac \partial {\partial \beta }\ln Z
\end{equation}
\begin{equation}
S=\beta \left\langle E\right\rangle +\ln Z.  \label{S}
\end{equation}
Eqs.$\left( \ref{Z}\right) $-$\left( \ref{S}\right) $ are the so-called
quantum gravitational statistic thermodynamics in canonical ensemble.

Now the problem is how to define the Euclidean action of the above manifold.

As is known, the Euclidean section $S^4$ of the de Sitter manifold has no
boundary, so the contribution to the action of the de Sitter space comes
only from the volume term. 
\begin{eqnarray}
I &=&\text{valume term}=\frac 1{16\pi }\int \sqrt{-g}\left( R-2\Lambda
\right) d^4x  \nonumber \\
&=&\frac 1{16\pi }\int \sqrt{-g}\left( R-2\Lambda \right) dtd^3x.  \label{I}
\end{eqnarray}
In order that the path-integral 
\begin{equation}
\int D\left[ g\right] \exp \left( iI\right)
\end{equation}
be convergent, we usually put 
\begin{equation}
-iI\equiv \hat{I},\quad \text{or }\int D\left[ g\right] \exp \left( -\hat{I}%
\right)  \label{Ea}
\end{equation}
wherein the wick rotation of the time variable $t\rightarrow -i\tau $ in $I$
is used, $\hat{I}$ is then the so-called Euclidean action. As is known from
positive action theorem, the Euclidean action $\hat{I}$ obtained by such way
is always positive for matter field and certain gravitational field (e.g.
Schwarzschild spacetime) but it is not for de Sitter space-time. As can be
easily shown from the above Euclideanization $\left( \ref{Ea}\right) $ of $%
\left( \ref{I}\right) $ \cite{G-H} 
\begin{equation}
\hat{I}=-\frac 1{16\pi }\int \sqrt{g}\left( R-2\Lambda \right) d\tau
d^3x=-3\pi \Lambda ^{-1}<0  \label{Ie}
\end{equation}
for Euclidean de Sitter space $\left( \ref{metric}\right) $. However now Eq.$%
\left( \ref{Ea}\right) $ can't be convergent for the ``wrong sign $-$'' in
Eq.$\left( \ref{Ie}\right) $. Instead of $\left( \ref{Ea}\right) $, if we
put 
\begin{equation}
-iI=\hat{I}
\end{equation}
but anti-Wick rotation 
\begin{equation}
t\rightarrow i\tau
\end{equation}
is used, then 
\begin{equation}
\hat{I}=\frac 1{16\pi }\int \sqrt{g}\left( R-2\Lambda \right) d\tau
d^3x=3\pi \Lambda ^{-1}>0.
\end{equation}
The ``wrong sign'' in Eq.$\left( \ref{Ie}\right) $ disappears and Eq.$\left( 
\ref{Ea}\right) $ becomes 
\begin{equation}
\int D\left[ g\right] \exp \left( -\hat{I}\right) ,\quad \hat{I}>0
\end{equation}
which is convergent.

We don't wish to be involved in the debate of choosing Wick rotation or
anti-Wick rotation in field theory including gravitation raised by Hawking
and Linde several years ago\cite{L-H}. Any way we believe that in de Sitter
case the only right choice seems to be anti-Wick rotation.

From Eqs.$\left( \ref{Z}\right) $ and $\left( \ref{S}\right) $, we can
easily get for de Sitter space 
\begin{eqnarray}
\hat{I} &=&\frac 1\pi \beta ^2, \\
\left\langle E\right\rangle &=&\frac 1{2\pi }\beta =\sqrt{\frac 3\Lambda },
\\
S &=&\frac 1{4\pi }\beta ^2=\frac{3\pi }\Lambda =\frac 14A_c,
\end{eqnarray}
where $A_c=12\pi \Lambda ^{-1}$ is the area of the cosmological horizon of
de Sitter space and the average energy $\left\langle E\right\rangle $ of the
de Sitter space in canonical ensemble is two times the classical vacuum
energy $E_{vac}=\left( 1/2\right) \sqrt{3/\Lambda }=\beta /\left( 2\pi
\right) $ of de Sitter space. The above calculation shows one can equally
well using partition function of the canonical ensemble to get the entropy
of de Sitter space if anti-Wick rotation is used.

\section{One loop quantum correction of the entropy of de Sitter space}

Hawking conjectured many years ago that the space-time foam may be formed
from $S^2\times S^2$ bubbles\cite{Hawking,Hawking2}. However a proof of this
conjecture is wanted. Recently in Ref.\cite{L-Hf} it is shown that the $%
S^2\times S^2$ bubbles can be created really from vacuum fluctuation in
one-loop approximation of both steady state universe 
\begin{equation}
ds^2=-dt^2+\exp \left( \frac{2t}\alpha \right) \sum_{i=1}^3\left(
dx^i\right) ^2,\quad \alpha \equiv \sqrt{\frac 3\Lambda }  \label{ssu}
\end{equation}
and closed de Sitter universe 
\begin{equation}
ds^2=-dt^2+\alpha ^2\cosh ^2\left( \frac t\alpha \right) \left( d\chi
^2+\sin ^2\chi d\Omega ^2\right) .  \label{cds}
\end{equation}
A brief account of the proof is given as follows.

The space-time metrics $\left( \ref{ssu}\right) $ and $\left( \ref{cds}%
\right) $ are conformal flat and their conformal vacuum is Minkovsky vacuum $%
\left| 0\right\rangle _M$. In such conformal trivial case the renormalized
vacuum energy-stress tensor from vacuum fluctuation of space-time $\left( 
\ref{ssu}\right) $ and $\left( \ref{cds}\right) $ in one-loop approximation
is given in signature $+2$ by\cite{Hu} 
\begin{equation}
\left\langle 0\left| T_\mu ^\nu \right| 0\right\rangle _{ren}=-\frac{q\left(
s\right) }{960\pi ^2\alpha ^4}g_\mu ^\nu ,\quad \alpha =\sqrt{\frac 3\Lambda 
},  \label{rnd}
\end{equation}
where $s$ is the spin of the matter field, 
\begin{equation}
q\left( 0\right) =1,q\left( \frac 12\right) =\frac{11}2,q\left( 1\right) =62.
\end{equation}
Henceforth we put $q\left( s\right) =q\left( 0\right) =1$.

For obtaining the foam structure of certain background spacetime, we should
consider the loop expansion of the effective action. In the semiclassical
Einstein gravitational theory, the loop expansion of the effective action $%
\Gamma _m\left[ g\right] $ of a quantum matter field can be expressed as\cite
{Hu} 
\begin{equation}
\Gamma _m\left[ g,\varphi \right] =\frac 1\hbar \Gamma _m^{\left( 0\right)
}\left[ g,\varphi \right] +\Gamma _m^{\left( 1\right) }\left[ g,\varphi
\right] +\hbar \Gamma _m^{\left( 2\right) }\left[ g,\varphi \right] +\cdot
\cdot \cdot ,  \label{mle}
\end{equation}
where $\hbar $ is the Planck constant, $\Gamma _m^{\left( 0\right) }\left[
g,\varphi \right] $ is the classical matter action or 0-loop approximation
of quantum matter field, $\Gamma _m^{\left( 1\right) }\left[ g,\varphi
\right] $ is the renormalized 1-loop contribution to the effective action of
quantum matter field, $\Gamma _m^{\left( 2\right) }\left[ g,\varphi \right] $
is the renormalized 2-loop contribution to the effective action of quantum
matter field, and so on.

The semiclassical Einstein gravitational field equation then reads 
\begin{equation}
\delta \left[ I_g\left[ g\right] +\Gamma _m\left[ g,\varphi \right] \right]
=0,  \label{SEQ}
\end{equation}
where $I_g\left[ g\right] $ is the classical Einstein-Hilbert action.
Substituting Eq.$\left( \ref{mle}\right) $ into Eq.$\left( \ref{SEQ}\right) $%
, one can obtain the so called self-consistent solution of Eq.$\left( \ref
{SEQ}\right) $, where the back reaction of quantum matter field on the
classical background geometry is included. However in this calculation no
separation of spacetime foam-like structure from the classical background
geometry structure is possible, so the idea of spacetime foam-like picture
is rather ambiguous. But if one notes that the deviation of the classical
background geometry caused by n-th-loop contribution of the quantum matter
field, say $\Gamma _m^{\left( 1\right) }\left[ g,\varphi \right] $, should
be the same order of magnitude as $\Gamma _m^{\left( 1\right) }\left[
g,\varphi \right] $ and so on, we may also give a similar ``loop-expansion''
of $I_g\left[ g\right] $ as 
\begin{equation}
I_g\left[ g\right] =\frac 1\hbar I_g^{\left( 0\right) }\left[ g\right]
+I_g^{\left( 1\right) }\left[ g\right] +\hbar I_g^{\left( 2\right) }\left[
g\right] +\cdot \cdot \cdot ,  \label{GLE}
\end{equation}
though the gravitational field now is not quantized. Inserting Eqs.$\left( 
\ref{mle}\right) $ and $\left( \ref{GLE}\right) $ into Eq.$\left( \ref{SEQ}%
\right) $, just by counting powers of $\hbar $, we can obtain a series of
gravitational field equation 
\begin{eqnarray}
\delta \left( I_g^{\left( 0\right) }+\Gamma _m^{\left( 0\right) }\right)
&=&0,  \label{A1} \\
\delta \left( I_g^{\left( 1\right) }+\Gamma _m^{\left( 1\right) }\right)
&=&0,  \label{A2} \\
\delta \left( I_g^{\left( 2\right) }+\Gamma _m^{\left( 2\right) }\right)
&=&0,  \label{A3} \\
&&\cdot \cdot \cdot \cdot \cdot \cdot ,  \nonumber
\end{eqnarray}
where the solution of equation $\left( \ref{A1}\right) $ is just the
classical background geometry. The solution of equation $\left( \ref{A2}%
\right) $ is the geometry created from 1-loop contribution of quantum matter
vacuum fluctuation, and so on.

The advantage of the above loop expansion Eqs.$\left( \ref{mle}\right) $ and 
$\left( \ref{GLE}\right) $ is that now we have a clear separation of the
background geometry from the spacetime foam geometry created from the
fluctuation of the vacuum matter field of the classical background geometry,
so as a result, a concrete picture of spacetime foam-like structure is
obtained.

According to Hawking's conjecture the spacetime foam-like structure might be
consisted of $S^2\times S^2$ bubble, which is a solution (Nariai solution)
of Einstein gravitational field equation, so we may further assume that all
``loop actions'' of different order are of Einstein-Hilbert type.

Now if the bubbles are really created from one-loop vacuum fluctuation of
the background space-time in signature $\left( +2\right) $, then Eq.$\left( 
\ref{A2}\right) $ should be 
\begin{equation}
G_\mu ^\nu =8\pi \left\langle 0\left| T_\mu ^\nu \right| 0\right\rangle
_{ren},  \label{GT}
\end{equation}
which is satisfied by the renormalizd vacuum energy-stress tensor $\left( 
\ref{rnd}\right) $ and the Einstein tensor $G_\mu ^\nu $ of the bubble $%
S^2\times S^2$.

As is known the metric of $S^2\times S^2$ is given by Nariai, Bousso and
Hawking\cite{B-H} 
\begin{equation}
ds^2=\frac 1\lambda \left( \sin ^2\chi d\Psi ^2+d\chi ^2+d\Omega ^2\right)
\label{Es1}
\end{equation}
in Euclidean signature, and 
\begin{equation}
ds^2=\frac 1\lambda \left( -\sin ^2\chi d\Psi ^2+d\chi ^2+d\Omega ^2\right)
\label{Es2}
\end{equation}
in Lorentzian signature of $\left( +2\right) $, where $\lambda ^{-1/2}$ is
the radius of the 2-sphere $S^2$.

It is straightforward to show that the Einstein tensor $G_\mu ^\nu $ of Eq.$%
\left( \ref{Es2}\right) $ is\cite{L-Hf} 
\begin{equation}
G_\mu ^\nu =-\lambda g_\mu ^\nu .
\end{equation}
So Eq.$\left( \ref{GT}\right) $ is satisfied if 
\begin{equation}
\lambda =\frac{8\pi }{9\times 960\pi ^2}\Lambda ^2=\frac 1{120\pi \alpha ^4}=%
\frac{\pi ^3}{120}\beta ^{-4},\quad \beta =\pi \alpha .  \label{lam}
\end{equation}

Now we would like to study the one-loop correction of the entropy of de
Sitter space from $S^2\times S^2$ bubbles.

For de Sitter space of metric $\left( \ref{metric}\right) $, $\left( \ref
{ssu}\right) $ and $\left( \ref{cds}\right) $, we have 
\begin{equation}
\hat{I}=3\pi \Lambda ^{-1}=\frac 1{4\pi }\beta ^2.
\end{equation}
However the vacuum energy-stress tensor $\left( \ref{rnd}\right) $ of vacuum
fluctuation in one-loop approximation is the same only for steady state de
Sitter space $\left( \ref{ssu}\right) $ and closed de Sitter space $\left( 
\ref{cds}\right) $, but different for static de Sitter space $\left( \ref
{metric}\right) $, the latter is\cite{B-D} 
\begin{eqnarray}
\left\langle 0\left| T_\mu ^\nu \right| 0\right\rangle _{static\;dS}
&=&\left\langle 0\left| T_\mu ^\nu \right| 0\right\rangle _{\left( \ref{rnd}%
\right) }+\left\langle 0\left| T_\mu ^\nu \right| 0\right\rangle _{Rindler} 
\nonumber \\
&=&-\frac{q\left( s\right) }{960\pi ^2\alpha ^4}\delta _\mu ^\nu -\frac{%
p\left( s\right) }{2\pi ^2}\left( \alpha ^2-r^2\right) ^{-2}diag.\left( 1,-%
\frac 13,-\frac 13,-\frac 13\right) ,
\end{eqnarray}
for the conformal vacuum of $\left( \ref{metric}\right) $ is Rindler vacuum.
Here $\left\langle 0\left| T_\mu ^\nu \right| 0\right\rangle _{\left( \ref
{rnd}\right) }$ means the renormalizied vacuum energy-stress tensor of Eq.$%
\left( \ref{rnd}\right) $. However their trace is the same, i.e., 
\begin{equation}
\left\langle 0\left| T_\mu ^\mu \right| 0\right\rangle
_{static\;dS}=\left\langle 0\left| T_\mu ^\mu \right| 0\right\rangle
_{\left( \ref{rnd}\right) }+0.
\end{equation}
So from $R=8\pi \left\langle T\right\rangle $ the scalar curvature of the
bubble created from the vacuum fluctuation of the background static de
Sitter space $\left( \ref{metric}\right) $ is the same as that from
background de Sitter spaces $\left( \ref{ssu}\right) $ and $\left( \ref{cds}%
\right) $. This implies that the action $\hat{I}_{one-loop}$ of $S^2\times
S^2$ bubble for the above three different de Sitter spaces $\left( \ref
{metric}\right) $, $\left( \ref{ssu}\right) $ and $\left( \ref{cds}\right) $
are the same. It is easy to find\cite{Hawking,B-H} 
\begin{equation}
\hat{I}_{one-loop}=\hat{I}\left[ S^2\times S^2\right] =2\pi \lambda ^{-1}=%
\frac{15}{\pi ^2}\beta ^4,
\end{equation}
where the anti-Wick rotation and the relation $\left( \ref{lam}\right) $ are
used.

Then the one-loop correction of the action and entropy of the de Sitter
space are 
\begin{equation}
\hat{I}=\frac 1{4\pi }\beta ^2+\frac{15}{\pi ^2}\beta ^4
\end{equation}
and 
\begin{equation}
S=-\beta \frac \partial {\partial \beta }\ln Z+\ln Z=\frac 1{4\pi }\beta ^2+%
\frac{45}{\pi ^2}\beta ^4  \label{sc}
\end{equation}
respectively.

Obviously, the above formulas only hold when $\beta ^4<<\beta ^2$, or $%
\Lambda >>L_p^{-2}\sim 10^{66}cm^{-2}$, where $L_p$ is the Planck length.
This is the case for very early inflating universe of size $\sqrt{3/\Lambda }%
<<L_p$. The results of $\left( \ref{sc}\right) $ implies, that if the de
Sitter space could be looked upon as a canonical ensemble, then the bubbles
created from vacuum fluctuation in 1-loop approximation will induce an
entropy increase of $45\beta ^4/\pi ^2$ to the de Sitter space. It is
reasonable, for such an increase of the entropy implies an increase of
disorder when vacuum fluctuation is considered.

\acknowledgments 

We thank Dr. Chao-Guang Huang for his helpful disscussion. This work is
supported by National Natural Sciences Foundation of China under grant
19473005.

\end{document}